\documentclass[11pt]{article}
\usepackage{amssymb,amsmath,amsfonts}

\begin{document}

\title{Generalized Abel-Plana formula as a renormalization tool in quantum
field theory with boundaries}
\author{A. A. Saharian\thanks{%
E-mail: saharyan@server.physdep.r.am} \\
%EndAName
\\
{\it Department of Physics, Yerevan State University,}\\
{\it 375025 Yerevan, Armenia}\\
{\it and}\\
{\it Departamento de F\'{\i}sica-CCEN, Universidade Federal da Para\'{\i}ba,}%
\\
{\it 58.059-970, Caixa Postal 5.008, Jo\~{a}o Pessoa, PB, Brazil}}
\maketitle

\begin{abstract}
We apply the generalized Abel-Plana formula for the investigation
of one-loop quantum effects on manifolds with boundaries. This
allows to extract from the vacuum expectation values of local
physical observables the parts corresponding to the geometry
without boundaries and to present the boundary-induced parts in
terms of integrals strongly convergent for the points away from
the boundaries. As a result, the renormalization procedure for
these observables is reduced to the corresponding procedure for
the bulks without boundaries.
\end{abstract}

\section{Introduction}

In many physical problems we need to consider the model on
background of manifolds with boundaries on which the dynamical
variables satisfy some prescribed boundary conditions. In quantum
field theory the imposition of boundary conditions leads to the
modification of the spectrum for the zero-point fluctuations and
results in the shift in the vacuum expectation values for physical
observables. In particular, vacuum forces arise acting on the
constraining boundary. This is the familiar Casimir effect (see
Refs. \cite{Grib94,Most97,Bord01}). Within the framework of the
mode summation method, in calculations of the expectation values
for physical observables one often needs to sum over the values of
a certain function at integer points, and then subtract the
corresponding quantity for unbounded space (usually presented in
terms of integrals). Practically, the sum and integral, taken
separately, diverge and some physically motivated procedure to
handle the finite result, is needed (for a discussion of different
methods to evaluate this finite difference see Refs.
\cite{Bart80}). For a number of geometries one of the most
convenient methods to obtain the renormalized values of the
mode-sums is based on the use of the Abel-Plana formula (APF)
\cite{Hardy}. In Ref. \cite{Mamaev} this formula has been used for
the investigation of the expectation values of the energy
momentum-tensor for a scalar field on cosmological backgrounds.
Further applications to the Casimir effect with corresponding
references can be found in \cite{Grib94,Most97}. The use of the
APF allows to extract in a manifestly cutoff independent way the
contribution of the unbounded space and present the renormalized
values in terms of exponentially converging integrals.

The applications of the APF in its usual form is restricted to the
problems where the eigenmodes have simple dependence on quantum
numbers and the normal modes are explicitly known. For the case of
curved boundaries and for mixed boundary conditions the normal
modes are given implicitly as zeroes of the corresponding
eigenfunctions or their combinations. To include more general
class of problems, in Ref. \cite{Sah1} the APF has been
generalized (see also Ref. \cite{Sahdis}). The generalized version
contains two meromorphic functions and the APF is obtained by
specifying one of them (for other generalizations of the APF see
\cite{Most97,Bart80}). Applying the generalized formula to Bessel
functions in Refs. \cite{Sah1, Sahdis} summation formulae are
obtained for the series over the zeros of various combinations of
these functions (for a review with physical applications see also
Ref. \cite{Saha00Rev}). In the present paper we give applications
of the generalized Abel-Plana formula to physical problems. The
paper is organized as follows. In the next section we briefly
outline the generalized Abel-Plana formula and discuss special
cases. In section \ref{sec:Sph} we consider the vacuum
polarization by spherical boundaries on background of global
monopole spacetime. The application of the generalized Abel-Plana
formula to the problem of the influence of uniformly accelerated
mirrors on
the properties of the Fulling-Rindler vacuum is given in section \ref%
{sec:Accel}. Section \ref{sec:Applic} describes some other
applications.

\section{Generalized Abel-Plana formula}

\label{sec:GAPF}

Let $f(z)$ and $g(z)$ be meromorphic functions for $a\leqslant
x\leqslant b$ in the complex plane $z=x+iy$. We denote by
$z_{f,k}$ and $z_{g,k}$ the poles of $f(z)$ and $g(z)$ in the
region $a<x<b$, respectively, and assume that ${\rm
Im\,}z_{f,k}\neq 0$. By using the residue theorem, it can be seen
\cite{Sah1} that if the functions $f(z)$ and $g(z)$ satisfy the
conditions
\begin{equation}
\lim_{h\rightarrow \infty }\int_{a\pm ih}^{b\pm ih}dz\,\left[ g(z)\pm f(z)%
\right] =0,\;\lim_{b\rightarrow \infty }\int_{b}^{b\pm i\infty
}dz\,\left[ g(z)\pm f(z)\right] =0,  \label{cor11}
\end{equation}%
and the integral $\int_{a}^{b}dx\,f(x)$ converges, then
\begin{equation}
\lim_{b\rightarrow \infty }\left\{
\int_{a}^{b}dx\,f(x)-R[f(z),g(z)]\right\}
=\frac{1}{2}\int_{a-i\infty }^{a+i\infty }dz\,\left[ g(z)+{\rm sgn}({\rm Im}%
z)f(z)\right] ,  \label{th12}
\end{equation}%
where
\begin{equation}
R[f(z),g(z)]=\pi i\left[ \sum_{k}\underset{z=z_{g,k}}{{\rm Res}}g(z)+\sum_{k}%
\underset{{\rm Im}z_{f,k}>0}{{\rm Res}}f(z)-\sum_{k}\underset{{\rm Im}%
z_{f,k}<0}{{\rm Res}}f(z)\right] .  \label{cor13}
\end{equation}%
We will refer to formula (\ref{th12}) as generalized Abel-Plana
formula (GAPF) as for $b=n+a$, $0<a<1$, $g(z)=-if(z)\cot \pi z$,
and for analytic functions $f(z)$ from (\ref{th12}) follows the
APF \cite{Hardy}
\begin{equation}
\lim_{n\rightarrow \infty }\left[ \sum_{1}^{n}f(s)-\int_{a}^{n+a}dx\,f(x)%
\right] =\frac{1}{2i}\int_{a}^{a-i\infty }dz\,f(z)(\cot \pi z-i)-\frac{1}{2i}%
\int_{a}^{a+i\infty }dz\,f(z)(\cot \pi z+i).  \label{apsf1}
\end{equation}%
The useful form of (\ref{apsf1}) may be obtained performing the limit $%
a\rightarrow 0$. By taking into account that the point $z=0$ is a
pole for
the integrands and therefore has to be avoided by arcs of the small circle $%
C_{\rho }$ on the right and performing $\rho \rightarrow 0$, one
obtains
\begin{equation}
\sum_{n=0}^{\infty }f(n)=\int_{0}^{\infty }dx\,f(x)+\frac{1}{2}%
f(0)+i\int_{0}^{\infty }dx\,\frac{f(ix)-f(-ix)}{e^{2\pi x}-1}.
\label{apsf2}
\end{equation}%
Note that now conditions (\ref{cor11}) are satisfied if
$\lim_{y\rightarrow \infty }e^{-2\pi |y|}|f(x+iy)|=0$, uniformly
in any finite interval of $x$. Formula (\ref{apsf2}) is the most
frequently used form of the APF in physical applications. Another
useful form \cite{Grib94} (in particular, for fermionic field
calculations) to sum over the values of an analytic function at
half of an odd integer points can be obtained from (\ref{apsf2}).

Formula (\ref{th12}) contains two meromorphic functions and is too
general. To obtain more specific consequences we have to specify
one of them. In applications we often need to sum over the values
of a certain function at the points being the zeros of the other
function. In order to obtain the summation formula for this type
of series, as a function $g(z)$ we take the function for which
these zeros are poles. Then the first term in the square brackets
of Eq. (\ref{cor13}) will give the corresponding series. This
choice should be made in a way to meet conditions (\ref{cor11}).
We will illustrate this on specific physical examples of quantum
field theory with boundaries.

\section{Vacuum densities for spherical boundaries}

\label{sec:Sph}

Consider a real scalar field $\varphi $ with curvature coupling parameter $%
\zeta $ on a $(D+1)$-dimensional spacetime region $M$ with static boundary $%
\partial M$. The corresponding field equation has the form
\begin{equation}
\left( \nabla _{i}\nabla ^{i}+m^{2}+\zeta R\right) \varphi
=0,\quad \label{mfieldeq}
\end{equation}%
where $R$ is the scalar curvature for the background spacetime,
$m$ is the mass for the field quanta, $\nabla _{i}$ is the
covariant derivative operator. We assume that on the boundary the
field satisfies the Robin boundary condition
\begin{equation}
\left( \tilde{A}+\tilde{B}n^{i}\nabla _{i}\right) \varphi
(x)=0,\;x\in
\partial M,  \label{mrobcond}
\end{equation}%
with constant coefficients $\tilde{A}$, $\tilde{B}$ and $n^{i}$
being the inward-pointing normal to the boundary. The vacuum
expectation values (VEVs) for physical quantities bilinear in the
field operator, can be found evaluating one of two-point
functions. Here we will consider the Wightman function, as it also
determines the response of particle detectors in a given state of
motion. This function can be determined from the mode-sum formula
\begin{equation}
W(x,x^{\prime })=\langle 0|\varphi (x)\varphi (x^{\prime
})|0\rangle =\sum_{\sigma }\varphi _{\sigma }(x)\varphi _{\sigma
}^{\ast }(x^{\prime }), \label{mfieldmodesum}
\end{equation}%
where $\left\{ \varphi _{\sigma }(x),\varphi _{\sigma }^{\ast
}(x^{\prime })\right\} $ is a complete orthonormal set of positive
and negative
frequency solutions to the field equation, specified by quantum numbers $%
\sigma $ and satisfying boundary condition (\ref{mrobcond}).

\subsection{Vacuum inside a spherical shell in the global
monopole spacetime}

As the first example we consider the region inside a spherical
shell with radius $a$ in the global monopole spacetime. In the
hyperspherical polar coordinates $(r,\vartheta ,\phi )\equiv
(r,\theta _{1},\theta _{2},\ldots \theta _{n},\phi )$, $n=D-2$,
the corresponding line element has the form
\begin{equation}
ds^{2}=dt^{2}-dr^{2}-\alpha ^{2}r^{2}d\Omega _{D}^{2},
\label{mmetric}
\end{equation}%
where $d\Omega _{D}^{2}$ is the line element on the surface of
unit sphere in $D$-dimensional Euclidean space, the parameter
$\alpha $ is smaller than unity and is related to the symmetry
breaking energy scale in the theory. The solid angle corresponding
to Eq. (\ref{mmetric}) is $\alpha ^{2}S_{D}$, with $S_{D}=2\pi
^{D/2}/\Gamma (D/2)$ being the total area of the surface of the
unit sphere in $D$-dimensional Euclidean space.

For the region inside the sphere the complete set of solutions to Eq. (\ref%
{mfieldeq}) is specified by the set of quantum numbers $\sigma
=(\lambda
,m_{k})$, where $m_{k}=(m_{0}\equiv l,m_{1},\ldots ,m_{n})$ and $%
m_{1},m_{2},\ldots ,m_{n}$ are integers such that $0\leqslant
m_{n-1}\leqslant m_{n-2}\leqslant \cdots \leqslant m_{1}\leq l$, $%
-m_{n-1}\leqslant m_{n}\leqslant m_{n-1}$. The corresponding
eigenfunctions have the form
\begin{equation}
\varphi _{\sigma }(x)=\beta _{\sigma }r^{-n/2}J_{\nu _{l}}(\lambda
r)Y(m_{k};\vartheta ,\phi )e^{-i\omega t},\,\,\omega
=\sqrt{\lambda ^{2}+m^{2}},\;l=0,1,2,\ldots ,  \label{meigfunc}
\end{equation}%
where $J_{\nu _{l}}(z)$ is the Bessel function of the order
\begin{equation}
\nu _{l}=\frac{1}{\alpha }\left[ \left( l+\frac{n}{2}\right)
^{2}+(1-\alpha ^{2})n\left( (n+1)\zeta -\frac{n}{4}\right) \right]
^{1/2},  \label{nuel}
\end{equation}%
and $Y(m_{k};\vartheta ,\phi )$ is the hyperspherical harmonic of degree $l$%
. The coefficients $\beta _{\sigma }$ can be found from the
normalization condition and is equal to
\begin{equation}
\beta _{\sigma }^{2}=\frac{\lambda T_{\nu _{l}}(\lambda
a)}{N(m_{k})\omega a\alpha ^{D-1}},\;T_{\nu }(z)\equiv
\frac{z}{(z^{2}-\nu ^{2})J_{\nu }^{2}(z)+z^{2}J_{\nu }^{^{\prime
}2}(z)}.  \label{normcoef}
\end{equation}

From boundary condition (\ref{mrobcond}) it follows that the
possible values for $\lambda $ are solutions to the equation
\begin{equation}
AJ_{\nu _{l}}(z)+BzJ_{\nu _{l}}^{\prime }(z)=0,\quad z=\lambda a,\quad A=%
\tilde{A}-nB/2,\quad B=-\tilde{B}/a.  \label{eigenmodes}
\end{equation}%
It is well known that for real $A$, $B$, and $\nu _{l}>-1$ all
roots of this equation are simple and real, except the case
$A/B<-\nu _{l}$ when there are two purely imaginary zeros. In the
following we will assume values of $A/B$ for which all roots are
real, $A/B\geq -\nu _{l}$. Let $\lambda _{\nu
_{l},k},\,k=1,2,\ldots ,$ be the positive zeros of the function
$AJ_{\nu _{l}}(z)+BzJ_{\nu _{l}}^{\prime }(z)$, arranged in
ascending order. The corresponding eigenfrequencies $\omega
=\omega _{\nu _{l},k}$ are related to these zeros by the formula
$\omega _{\nu _{l},k}=\sqrt{\lambda _{\nu
_{l},k}^{2}/a^{2}+m^{2}}$. Substituting Eq. (\ref{meigfunc}) into Eq. (\ref%
{mfieldmodesum}) and using the addition formula for the spherical
harmonics, one obtains
\begin{eqnarray}
W(x,x^{\prime })&=&\frac{(rr^{\prime })^{-n/2}}{naS_{D}\alpha ^{D-1}}%
\sum_{l=0}^{\infty }(2l+n)C_{l}^{n/2}(\cos \theta ) \nonumber \\
&& \times \sum_{k=1}^{\infty }\frac{%
\lambda _{\nu _{l},k}T_{\nu _{l}}(\lambda _{\nu
_{l},k})}{\sqrt{\lambda _{\nu _{l},k}^{2}+m^{2}a^{2}}}J_{\nu
_{l}}(\lambda _{\nu _{l},k}r/a)J_{\nu _{l}}(\lambda _{\nu
_{l},k}r^{\prime }/a)e^{i\omega _{\nu _{l},k}(t^{\prime }-t)},
\label{fieldmodesum1}
\end{eqnarray}%
where $C_{p}^{q}(x)$ is the Gegenbauer polynomial of degree $p$ and order $q$%
, and $\theta $ is the angle between the directions $(\vartheta
,\phi )$ and $(\vartheta ^{\prime },\phi ^{\prime })$. As the
normal modes $\lambda _{\nu
_{l},k}$ are not explicitly known, the Wightman function in the form (\ref%
{fieldmodesum1}) is not convenient for the evaluation of the VEVs
for the
observables bilinear in the field. In addition, the terms with large values $%
k$ are highly oscillatory. In order to obtain summation formula
for the series over $k$, as a function $g(z)$ in the GAPF we
choose $g(z)=i\bar{Y}_{\nu }(z)f(z)/\bar{J}_{\nu }(z)$, where
$Y_{\nu }(z)$ is the Neumann function and for a given function
$F(z)$ we use the barred notation
\begin{equation}
\bar{F}(z)\equiv AF(z)+BzF^{\prime }(z).  \label{barnot}
\end{equation}%
The conditions (\ref{cor11}) are satisfied if the function $f(z)$
is restricted by the constraint
\begin{equation}
|f(z)|<\epsilon (x)e^{c|y|},\;z=x+iy,\quad |z|\rightarrow \infty ,
\label{condf}
\end{equation}%
where $c<2$ and $\epsilon (x)\rightarrow 0$ for $x\rightarrow
\infty $.
Assuming that the function $f(z)$ is analytic in the right half-plane, for (%
\ref{cor13}) one finds $R[f(z),g(z)]=2\sum_{k}T_{\nu }(\lambda
_{\nu ,k})f(\lambda _{\nu ,k})$. Substituting this expression into
(\ref{th12}) and taking the limit $a\rightarrow 0$, it can be seen
that the following summation formula takes place
\begin{eqnarray}
\sum_{k=1}^{\infty }T_{\nu }(\lambda _{\nu ,k})f(\lambda _{\nu ,k}) &=&\frac{%
1}{2}\int_{0}^{\infty }dx \, f(x)+{}\frac{\pi }{4}\underset{z=0}{{\rm Res}}f(z)%
\frac{\bar{Y}_{\nu }(z)}{\bar{J}_{\nu }(z)}  \nonumber \\
&&-\frac{1}{2\pi }\int_{0}^{\infty }dx \, \frac{\bar{K}_{\nu
}(x)}{\bar{I}_{\nu }(x)}\left[ e^{-\nu \pi i}f(ix)+e^{\nu \pi
i}f(-ix)\right] ,  \label{sumJ1}
\end{eqnarray}%
where we have introduced the modified Bessel functions $I_{\nu }(x)$ and $%
K_{\nu }(x)$. By taking in this formula $\nu =1/2$, $A=1$, $B=0$,
as a
special case we receive the APF in the form (\ref{apsf2}). Formula (\ref%
{sumJ1}) can be generalized in the case of the existence of purely
imaginary zeros for the function $\bar{J}_{\nu }(z)$ by adding the
corresponding residue term and taking the principal value of the
integral on the right (see Ref. \cite{Saha00Rev}).

For the summation over $k$ in Eq. (\ref{fieldmodesum1}) we apply formula (%
\ref{sumJ1}). The corresponding conditions are satisfied if
$r+r^{\prime }+|t-t^{\prime }|<2a$. In particular, the latter
constraint takes place in the coincidence limit for the points
away from the boundary. Now the Wightman function is presented in
the form
\begin{equation}
W(x,x^{\prime })=W_{{\rm m}}(x,x^{\prime })+\langle \varphi
(x)\varphi (x^{\prime })\rangle _{b},  \label{unregWight1}
\end{equation}%
where the term
\begin{equation}
W_{{\rm m}}(x,x^{\prime })=\frac{\alpha ^{1-D}}{2nS_{D}}\sum_{l=0}^{\infty }%
\frac{2l+n}{(rr^{\prime })^{n/2}}C_{l}^{n/2}(\cos \theta
)\int_{0}^{\infty
}dz\,\frac{ze^{i\sqrt{z^{2}+m^{2}}(t^{\prime }-t)}}{\sqrt{z^{2}+m^{2}}}%
J_{\nu _{l}}(zr)J_{\nu _{l}}(zr^{\prime }),  \label{Mink}
\end{equation}%
comes from the first integral on the right of Eq. (\ref{sumJ1})
and
\begin{eqnarray}
\langle \varphi (x)\varphi (x^{\prime })\rangle _{b} &=&-\frac{\alpha ^{1-D}%
}{\pi nS_{D}}\sum_{l=0}^{\infty }\frac{2l+n}{(rr^{\prime })^{n/2}}%
C_{l}^{n/2}(\cos \theta )\int_{m}^{\infty }dz\,z\frac{\bar{K}_{\nu _{l}}(az)%
}{\bar{I}_{\nu _{l}}(az)}  \label{Wightbound} \\
&&\times \frac{I_{\nu _{l}}(zr)I_{\nu _{l}}(zr^{\prime })}{\sqrt{z^{2}-m^{2}}%
}\cosh \left[ \sqrt{z^{2}-m^{2}}(t^{\prime }-t)\right] . \nonumber
\end{eqnarray}%
The part (\ref{Mink}) in the Wightman function does not depend on
$a$,
whereas the contribution of the term (\ref{Wightbound}) tends to zero as $%
a\rightarrow \infty $. It follows from here that expression
(\ref{Mink}) is the Wightman function for the unbounded global
monopole spacetime. This can be seen also by explicit evaluation
of the mode-sum using the eigenfunctions for the global monopole
spacetime without boundaries.

Having the Wightman function we can evaluate the VEVs for the
field square
and the energy-momentum tensor taking the following coincidence limits:%
\begin{eqnarray}
\langle 0|\varphi ^{2}|0\rangle &=&\lim_{x^{\prime }\rightarrow
x}W(x,x^{\prime }),  \label{mphi2VEV} \\
\langle 0|T_{ik}|0\rangle &=&\lim_{x^{\prime }\rightarrow
x}\partial
_{i}\partial _{k}^{\prime }W(x,x^{\prime })+\left[ \left( \zeta -\frac{1}{4}%
\right) g_{ik}\nabla _{l}\nabla ^{l}-\zeta \nabla _{i}\nabla
_{k}-\zeta R_{ik}\right] \langle 0|\varphi ^{2}|0\rangle .
\label{mTikVEV}
\end{eqnarray}%
Similar to the Wightman function, these VEVs are presented as sums
of the
boundary-free and boundary-induced parts and are investigated in Ref. \cite%
{Saha03}. Hence, the application of the GAPF allowed to extract
from the VEVs the parts corresponding to the background without
boundaries. For points away from the boundaries, the
boundary-induced parts are finite and the renormalization
procedure is needed for the boundary-free part only. This can be
done by using the standard methods of quantum field theory without
boundaries. In addition, the boundary-induced parts are presented
in terms of exponentially convergent integrals convenient for
numerical calculations.

For $\alpha =1$ the bulk corresponds to the $(D+1)$-dimensional
Minkowskian spacetime and one has $\nu _{l}=l+n/2$. In this case
the vacuum densities for a scalar field with Robin boundary
conditions on spherical boundaries are investigated in Ref.
\cite{Saha01}. In the case of the electromagnetic field on the
$D=3$ Minkowski bulk we have two types of modes. For the first one
(TE modes) the eigenmodes are zeroes of the function
$J_{l+1/2}(z)$, and for the second one the eigenmodes are
determined by the zeroes of the function $\bar{J}_{l+1/2}(z)$ with
$A/B=1/2$. On the base of the GAPF the corresponding VEVs for the
energy-momentum tensor are investigated in Ref. \cite{Grig86}.

\subsection{Region between two spheres}

In this section we are interested in the VEVs of the field
bilinear products on background of the geometry described by Eq.
(\ref{mmetric}), assuming that the field satisfies the Robin
boundary condition (\ref{mrobcond}) on two spheres with radii $a$
and $b$, $a<b$, concentric with the monopole. We will consider the
general case when the coefficients in the boundary conditions for
the inner and outer spheres are different and will denote them by
$\tilde{A}_{j}$ and $\tilde{B}_{j}$ with $j=a,b$. The
corresponding
eigenfunctions are given by the formula which is obtained from (\ref%
{meigfunc}) by the replacement of the radial part:%
\begin{equation}
J_{\nu _{l}}(\lambda r)\rightarrow g_{\nu _{l}}(\lambda a,\lambda
r)\equiv J_{\nu _{l}}(\lambda r)\bar{Y}_{\nu _{l}}^{(a)}(\lambda
a)-\bar{J}_{\nu _{l}}^{(a)}(\lambda a)Y_{\nu _{l}}(\lambda r),
\label{Jtog}
\end{equation}%
and the barred functions are defined in accordance with
\begin{equation}
\bar{F}^{(j)}(z)\equiv A_{j}F(z)+B_{j}zF^{\prime }(z),\quad A_{\alpha }=%
\tilde{A}_{j}-B_{j}n/2,\quad B_{j}=n^{(j)}\tilde{B}_{j}/j,\;j=a,b,
\label{barnotab}
\end{equation}%
where $n^{(a)}=-1$ and $n^{(b)}=1$. From the normalization
condition, for the coefficient in the eigenfunctions one finds
\begin{equation}
\beta _{\sigma }^{2}=\frac{\pi ^{2}\lambda T_{\nu _{l}}^{ab}(b/a,\lambda a)}{%
4N(m_{k})\omega a\alpha ^{D-1}},  \label{betalf2sph}
\end{equation}%
where we use the notation
\begin{equation}
T_{\nu }^{ab}(\eta ,z)=z\left\{ \frac{\bar{J}_{\nu
}^{(a)2}(z)}{\bar{J}_{\nu }^{(b)2}(\eta z)}\left[
A_{b}^{2}+B_{b}^{2}(\eta ^{2}z^{2}-\nu ^{2})\right]
-A_{a}^{2}-B_{a}^{2}(z^{2}-\nu ^{2})\right\} ^{-1},\quad \eta
=\frac{b}{a}. \label{tekaAB}
\end{equation}%
The functions chosen in the form (\ref{Jtog}) satisfy the boundary
condition on the sphere $r=a$. From the boundary condition on
$r=b$ one obtains that the corresponding eigenvalues for the
quantum number $\lambda $ are solutions to the equation
\begin{equation}
C_{\nu _{l}}^{ab}(b/a,\lambda a)\equiv \bar{J}_{\nu _{l}}^{(a)}(\lambda a)%
\bar{Y}_{\nu _{l}}^{(b)}(\lambda b)-\bar{J}_{\nu _{l}}^{(b)}(\lambda b)\bar{Y%
}_{\nu _{l}}^{(a)}(\lambda a)=0.  \label{eigmodesab}
\end{equation}%
Below the roots to this equation will be denoted by $\lambda
a=\gamma _{\nu _{l},k}$, $k=1,2,\ldots $.

Substituting the eigenfunctions into the mode-sum formula, one
finds
\begin{equation}
W(x,x^{\prime })=\frac{\pi ^{2}(rr^{\prime })^{-n/2}}{4naS_{D}\alpha ^{D-1}}%
\sum_{l=0}^{\infty }(2l+n)C_{l}^{n/2}(\cos \theta
)\sum_{k=1}^{\infty }h(\gamma _{\nu _{l},k})T_{\nu
_{l}}^{ab}(b/a,\gamma _{\nu _{l},k}), \label{fieldmodesum1ab}
\end{equation}%
with the function
\begin{equation}
h(z)=\frac{ze^{i\sqrt{z^{2}/a^{2}+m^{2}}(t^{\prime }-t)}}{\sqrt{%
z^{2}+m^{2}a^{2}}}g_{\nu _{l}}(z,zr/a)g_{\nu _{l}}(z,zr^{\prime
}/a). \label{hab}
\end{equation}%
To obtain a summation formula for the series over zeros of the function $%
C_{\nu }^{ab}(\eta ,z)$, in the GAPF as functions $g(z)$ and
$f(z)$ we take
\begin{equation}
g(z)=\frac{1}{2i}\left[ \frac{\bar{H}_{\nu }^{(1b)}(\eta
z)}{\bar{H}_{\nu
}^{(1a)}(z)}+\frac{\bar{H}_{\nu }^{(2b)}(\eta z)}{\bar{H}_{\nu }^{(2a)}(z)}%
\right] \frac{F(z)}{C_{\nu }^{ab}(\eta ,z)},\quad f(z)=\frac{F(z)}{\bar{H}%
_{\nu }^{(1a)}(z)\bar{H}_{\nu }^{(2a)}(z)},  \label{gefcomb}
\end{equation}%
where $F(z)$ is an analytic function in the right half-plane,
$H_{\nu
}^{(1,2)}(z)$ are the Hankel functions, and the notations $\bar{F}%
^{(j)},j=a,b$ are introduced in accordance with (\ref{barnotab}).
For the combinations entering in the right hand side of the GAPF
we obtain
\begin{equation}
g(z)-(-1)^{q}f(z)=-i\frac{\bar{H}_{\nu }^{(qa)}(\lambda
z)}{\bar{H}_{\nu }^{(qa)}(z)}\frac{F(z)}{C_{\nu }^{ab}(\eta
,z)},\quad q=1,2, \label{gefsumnew}
\end{equation}%
and for the residue term one has
\begin{equation}
\underset{z=\gamma _{\nu ,k}}{{\rm Res}}g(z)=\frac{\pi }{2i}T_{\nu
}^{ab}(\eta ,\gamma _{\nu ,k})F(\gamma _{\nu ,k}).  \label{rel31}
\end{equation}%
The conditions for the GAPF, written in terms of the function
$F(z)$, are as follows
\begin{equation}
|F(z)|<\epsilon _{1}(x)e^{c_{1}|y|},\quad |z|\rightarrow \infty
,\quad z=x+iy,  \label{cond31}
\end{equation}%
where $c_{1}<2(\eta -1)$, $x^{\delta _{B_{a}0}+\delta
_{B_{b}0}-1}\epsilon _{1}(x)\rightarrow 0$ for $x\rightarrow
+\infty $. From Eq. (\ref{th12}) we obtain the following summation
formula
\begin{eqnarray}
\sum_{k=1}^{\infty }F(\gamma _{\nu ,k})T_{\nu }^{ab}(\eta ,\gamma
_{\nu ,k}) &=&\frac{2}{\pi ^{2}}\int_{0}^{\infty
}\frac{F(x)dx}{\bar{J}_{\nu
}^{(a)2}(x)+\bar{Y}_{\nu }^{(a)2}(x)}-\frac{1}{\pi }\underset{z=0}{{\rm Res}}%
\left[ \frac{F(z)\bar{H}_{\nu }^{(1b)}(\eta z)}{C_{\nu }^{ab}(\eta ,z)\bar{H}%
_{\nu }^{(1a)}(z)}\right]   \nonumber \\
&&-\frac{1}{2\pi }\int_{0}^{\infty }dx\,\Omega _{a\nu }(x,\eta
x)\left[ F(ix)+F(-ix)\right] .  \label{cor3form}
\end{eqnarray}%
In Eq. (\ref{cor3form}) we have introduced the notation
\begin{equation}
\Omega _{a\nu }(x,\eta x)=\frac{\bar{K}_{\nu }^{(b)}(\eta
x)/\bar{K}_{\nu }^{(a)}(x)}{\bar{K}_{\nu }^{(a)}(x)\bar{I}_{\nu
}^{(b)}(\eta x)-\bar{K}_{\nu }^{(b)}(\eta x)\bar{I}_{\nu
}^{(a)}(x)}.  \label{Omega}
\end{equation}%
Note that (\ref{cor3form}) may be generalized for the functions
$F(z)$ having poles and in the case of the existence of purely
imaginary zeros for $C_{\nu }^{ab}(\eta ,z)$ \cite{Saha00Rev}.

Now, to sum the series over $k$ in Eq. (\ref{fieldmodesum1ab}), we take in (%
\ref{cor3form}) $F(z)=h(z)$. The corresponding conditions are satisfied if $%
r+r^{\prime }+|t-t^{\prime }|<2b$. In particular, this is the case
in the coincidence limit for the region between two spheres. As a
result, for the Wightman function one obtains
\begin{eqnarray}
W(x,x^{\prime }) &=&\frac{\alpha ^{1-D}}{2naS_{D}}\sum_{l=0}^{\infty }\frac{%
2l+n}{(rr^{\prime })^{n/2}}C_{l}^{n/2}(\cos \theta )\left\{
\int_{0}^{\infty
}\frac{h(z)dz}{\bar{J}_{\nu _{l}}^{(a)2}(z)+\bar{Y}_{\nu _{l}}^{(a)2}(z)}-%
\frac{2}{\pi }\int_{ma}^{\infty }dz\,z\right.   \nonumber \\
&&\times \left. \frac{\Omega _{a\nu }(z,\eta z)}{\sqrt{z^{2}-a^{2}m^{2}}}%
G_{\nu _{l}}^{(a)}(z,zr/a)G_{\nu _{l}}^{(a)}(z,zr^{\prime
}/a)\cosh \left[ \sqrt{z^{2}/a^{2}-m^{2}}(t^{\prime }-t)\right]
\right\} , \label{unregWightab}
\end{eqnarray}%
where we have introduced the notation
\begin{equation}
G_{\nu }^{(j)}(z,y)=I_{\nu }(y)\bar{K}_{\nu
}^{(j)}(z)-\bar{I}_{\nu }^{(j)}(z)K_{\nu }(y),\;j=a,b.
\label{Geab}
\end{equation}%
In the limit $b\rightarrow \infty $ the second integral on the right of (\ref%
{unregWightab}) tends to zero, whereas the first one does not
depend on $b$. It follows from here that the term with the first
integral in the figure braces corresponds to the Wightman function
for the region outside a single sphere with radius $a$ on
background of the global monopole geometry. The latter we will
denote by $W^{(a)}(x,x^{\prime })$. To extract from this function
the part induced by the presence of the sphere we use the identity
\begin{equation}
\frac{g_{\nu }(z,zr/a)g_{\nu }(z,zr^{\prime }/a)}{\bar{J}_{\nu }^{2}(z)+\bar{%
Y}_{\nu }^{2}(z)}=J_{\nu }(zr/a)J_{\nu }(zr^{\prime }/a)-\frac{1}{2}%
\sum_{s=1}^{2}\frac{\bar{J}_{\nu }(z)}{\bar{H}_{\nu
}^{(s)}(z)}H_{\nu }^{(s)}(zr/a)H_{\nu }^{(s)}(zr^{\prime }/a).
\label{relab}
\end{equation}%
The contribution from the first term on the right of this relation
gives the Wightman function $W_{{\rm m}}(x,x^{\prime })$ for the
geometry without boundaries. In the part coming from the second
term we rotate the contour for the integration over $z$ by the
angle $\pi /2$ for $s=1$ and by the angle $-\pi /2$ for $s=2$.
Introducing the modified Bessel functions, one finds
\begin{eqnarray}
W^{(a)}(x,x^{\prime }) &=&W_{{\rm m}}(x,x^{\prime })-\frac{\alpha ^{1-D}}{%
\pi nS_{D}}\sum_{l=0}^{\infty }\frac{2l+n}{(rr^{\prime })^{n/2}}%
C_{l}^{n/2}(\cos \theta )\int_{m}^{\infty }dz\,z  \nonumber \\
&&\times \frac{\bar{I}_{\nu _{l}}(az)}{\bar{K}_{\nu
_{l}}(az)}\frac{K_{\nu
_{l}}(zr)K_{\nu _{l}}(zr^{\prime })}{\sqrt{z^{2}-m^{2}}}\cosh \!\left[ \sqrt{%
z^{2}-m^{2}}(t^{\prime }-t)\right] .  \label{regWightout}
\end{eqnarray}%
As a result the Wightman function in the region between two
spheres is presented in the form
\begin{eqnarray}
W(x,x^{\prime }) &=&W^{(a)}(x,x^{\prime })-\frac{\alpha ^{1-D}}{\pi nS_{D}}%
\sum_{l=0}^{\infty }\frac{2l+n}{(rr^{\prime
})^{n/2}}C_{l}^{n/2}(\cos \theta )\int_{m}^{\infty
}dz\,z\frac{\Omega _{a\nu _{l}}(az,bz)}{\sqrt{z^{2}-m^{2}}}
\nonumber \\
&&\times G_{\nu _{l}}^{(a)}(az,rz)G_{\nu _{l}}^{(a)}(az,r^{\prime
}z)\cosh \left[ \sqrt{z^{2}-m^{2}}(t^{\prime }-t)\right] .
\label{regWightab1}
\end{eqnarray}%
In the coincidence limit, $x^{\prime }=x$, the second summand on
the right hand side of (\ref{regWightab1}) will give a finite
result for $a\leqslant r<b$, and is divergent on the boundary
$r=b$. Hence, in the problem with two boundaries, the GAPF allowed
to extract the part corresponding to the geometry with a single
boundary and to present the part induced by the second boundary in
terms of exponentially convergent integrals. The VEVs for the
field square and the energy-momentum tensor in the region between
two
spheres are obtained from the Wightman function on the base of formulae (\ref%
{mphi2VEV}), (\ref{mTikVEV}) and are investigated in Ref.
\cite{Saha04a}.

We have considered the case of a scalar field. The VEVs for the
energy-momentum tensor of a fermionic field $\psi $\ with the mass
$m$ on background of the global monopole have been discussed in
Ref. \cite{Saha04} for the case of a single spherical boundary and
in Ref. \cite{Beze06} for the geometry of two spherical boundaries
assuming that the field satisfies bag boundary condition $\left(
1+i\gamma ^{l}n_{l}\right) \psi =0$ on bounding surfaces. For the
region inside a spherical shell with radius $a$ the corresponding
eigenmodes are the zeroes of the function $\tilde{J}_{\nu _{\sigma
}}^{(a)}(z)$, where for a given function $F(z)$ we use the
notation
\begin{equation}
\tilde{F}^{(w)}(z)\equiv zF^{\prime }(z)+\left[ n^{(w)}\left( mw-\sqrt{%
z^{2}+m^{2}w^{2}}\right) -(-1)^{\sigma }\nu _{\sigma }\right]
F(z), \label{tildenot}
\end{equation}%
with $n^{(a)}=1$, $\nu _{\sigma }=(j+1/2)/\alpha -(-1)^{\sigma
}/2$, $j$ is the total angular momentum, and $\sigma =0,1$
correspond to two types of eigenfunctions with different parities.
By using the GAPF, a summation formula for the series over zeroes
of the function $\tilde{J}_{\nu _{\sigma }}^{(a)}(z)$ is derived
in Ref. \cite{Saha04}. In the region between two concentric
spherical boundaries with radii $a$ and $b$, $a>b$, on the global
monopole bulk, the eigenmodes of the fermionic field are the
zeroes of the function $\tilde{J}_{\nu _{\sigma
}}^{(b)}(z)\tilde{Y}_{\nu _{\sigma }}^{(a)}(za/b)-\tilde{Y}_{\nu
_{\sigma }}^{(b)}(z)\tilde{J}_{\nu _{\sigma }}^{(a)}(za/b)$ with
$n^{(b)}=-1$. A summation formula for the corresponding series is
obtained in Ref. \cite{Beze06} and has been applied to the
investigation of the corresponding VEVs. The properties of the
electromagnetic vacuum in the region between two concentric
spheres in the Minkowski bulk are studied in Ref. \cite{Saha87sph}
by using the GAPF.

\section{Vacuum polarization by uniformly accelerated mirrors}

\label{sec:Accel}

It is well known that the uniqueness of the vacuum state is lost
when we work within the framework of quantum field theory in a
general curved spacetime or in non--inertial frames. For instance,
the vacuum state for a uniformly accelerated observer, the
Fulling-Rindler vacuum, turns out to be inequivalent to that for
an inertial observer. An interesting topic in the investigations
of the Casimir effect is the dependence of the vacuum
characteristics on the type of the vacuum. In this section we will
consider the application of the GAPF for the investigation of the
scalar vacuum polarization brought about by the presence of
infinite plane boundaries moving by uniform proper acceleration
through the Fulling-Rindler vacuum.

\subsection{Wightman function for a single plate}

Consider a massive scalar field with general curvature coupling
and satisfying Robin boundary condition on an infinite plane
moving with uniform proper acceleration. In the accelerated frame
it is convenient to use the
Rinlder coordinates with the line element%
\begin{equation}
ds^{2}=\xi ^{2}d\tau ^{2}-d\xi ^{2}-d{\bf x}^{2}.
\label{metricRin}
\end{equation}%
A world-line defined by $\xi ,{\bf x}={\rm const}$ describes an
observer with constant proper acceleration $\xi ^{-1}$. We will
assume that the plate is located at $\xi =a$ and will consider the
region on the right from the boundary, $\xi \geqslant a$, where
$a^{-1}$ is the proper acceleration of the plate. In the Rindler
coordinates, boundary condition (\ref{mrobcond}) takes the form
\begin{equation}
\left( \tilde{A}+\tilde{B}\partial /\partial \xi \right) \varphi
(x)=0,\quad \xi =a.  \label{boundRind}
\end{equation}%
For $\xi \geqslant a$ a complete set of solutions that are of
positive frequency with respect to $\partial /\partial \tau $ and
bounded as $\xi \rightarrow \infty $ is
\begin{equation}
\varphi _{\sigma }(x)=C_{\sigma }K_{i\omega }(\lambda \xi )e^{i{\bf kx}%
-i\omega \tau },\quad \sigma =(\omega ,{\bf k}),\;\lambda =\sqrt{k^{2}+m^{2}}%
.  \label{sol2}
\end{equation}%
From boundary condition (\ref{boundRind}) we find that the
possible values for $\omega $ have to be zeros of the function
$\bar{K}_{i\omega }(\lambda a) $, where the barred notation is
defined by (\ref{barnot}) with the
coefficients $A=\tilde{A},\;B=\tilde{B}/a$. We will denote these zeros by $%
\omega =\omega _{n}=\omega _{n}(k)$, $n=1,2,...$ arranged in
ascending
order, $\omega _{n}<\omega _{n+1}$. The coefficient $C_{\sigma }$ in (\ref%
{sol2}) is determined by the normalization condition:

\begin{equation}
C_{\sigma }^{2}=\frac{1}{(2\pi )^{D-1}}\frac{\bar{I}_{i\omega
_{n}}(\lambda a)}{\frac{\partial }{\partial \omega
}\bar{K}_{i\omega }(\lambda a)\mid _{\omega =\omega _{n}}}.
\label{normc}
\end{equation}

Substituting the eigenfunctions (\ref{sol2}) into the mode-sum formula (\ref%
{mfieldmodesum}), we obtain
\begin{equation}
W(x,x^{\prime })=\int \frac{d{}{\bf k}}{(2\pi )^{D-1}}\,e^{i{}{\bf k}({\bf x}%
-{\bf x}^{\prime })}\sum_{n=1}^{\infty }\frac{\bar{I}_{i\omega }(\lambda a)}{%
\frac{\partial }{\partial \omega }\bar{K}_{i\omega }(\lambda
a)}K_{i\omega }(\lambda \xi )K_{i\omega }(\lambda \xi ^{\prime
})e^{-i\omega (\tau -\tau ^{\prime })}|_{\omega =\omega _{n}}.
\label{emtdiag}
\end{equation}%
A summation formula for the series over zeros $\omega _{n}$ can be
obtained from formula (\ref{th12}) taking
\begin{equation}
f(z)=\frac{2i}{\pi }F(z)\sinh \pi z,\quad g(z)=\frac{\bar{I}_{iz}(\eta )+%
\bar{I}_{-iz}(\eta )}{\bar{K}_{iz}(\eta )}F(z),  \label{fgK}
\end{equation}%
with a function $F(z)$ analytic in the right half-plane. By using
the asymptotic formulae for the modified Bessel functions for
large values of the index, the conditions for the GAPF can be
written in terms of the function $F(z)$ as follows:
\begin{equation}
|F(z)|<\epsilon (|z|)e^{-\pi x}\left( |z|/\eta \right)
^{2|y|},\quad z=x+iy,\quad x>0,\quad |z|\rightarrow \infty ,
\label{condFKi}
\end{equation}%
where $|z|\epsilon (|z|)\rightarrow 0$ for $|z|\rightarrow \infty
$. From the GAPF we obtain the summation formula
\begin{eqnarray}
\sum_{n=1}^{\infty }\left. \frac{\bar{I}_{iz}(\eta )F(z)}{\partial \bar{K}%
_{iz}(\eta )/\partial z}\right| _{z=\omega _{n}} &=&\frac{1}{\pi ^{2}}%
\int_{0}^{\infty }dxF(x)\sinh \pi x-\frac{F_{0}\bar{I}_{0}(\eta )}{2\bar{K}%
_{0}(\eta )}  \nonumber \\
&&-\frac{1}{2\pi }\int_{0}^{\infty }dx\frac{\bar{I}_{x}(\eta )}{\bar{K}%
_{x}(\eta )}\left[ F(ix)+F(-ix)\right] ,  \label{sumformKi2}
\end{eqnarray}%
where $F_{0}=\lim_{z\rightarrow 0}zF(z)$.

Now for the summation over $n$ in formula (\ref{emtdiag}) we choose $%
F(z)=K_{iz}(\lambda \xi )K_{iz}(\lambda \xi ^{\prime })e^{-iz(\tau
-\tau ^{\prime })}$. It can be seen that condition (\ref{condFKi}) is satisfied if $%
a^{2}e^{|\tau -\tau ^{\prime }|}<|\xi \xi ^{\prime }|$. Note that
this is the case in the coincidence limit $\tau =\tau ^{\prime }$
for the points in the region under consideration, $\xi ,\xi
^{\prime }>a$. The contribution
coming from the first integral term on the right of formula (\ref{sumformKi2}%
) corresponds to the Wightman function for the Fulling-Rindler
vacuum without boundaries:
\begin{equation}
W_{{\rm R}}(x,x^{\prime })=\frac{1}{\pi ^{2}}\int d{}{\bf k}\,\frac{e^{i{}%
{\bf k}({\bf x}-{\bf x}^{\prime })}}{(2\pi
)^{D-1}}\int_{0}^{\infty }d\omega \sinh (\pi \omega )e^{-i\omega
(\tau -\tau ^{\prime })}K_{i\omega }(\lambda \xi )K_{i\omega
}(\lambda \xi ^{\prime }).  \label{emtRindler}
\end{equation}%
As a result, the Wightman function is presented in the form%
\begin{equation}
W(x,x^{\prime })=W_{{\rm R}}(x,x^{\prime })+\langle \varphi
(x)\varphi (x^{\prime })\rangle _{{\rm b}},  \label{WRb}
\end{equation}%
where the second term on the right is induced by the presence of
the plate:
\begin{equation}
\langle \varphi (x)\varphi (x^{\prime })\rangle _{{\rm b}}=-\frac{1}{\pi }%
\int d{}{\bf k}\,\frac{e^{i{}{\bf k}({\bf x}-{\bf x}^{\prime
})}}{(2\pi
)^{D-1}}\int_{0}^{\infty }d\omega \frac{\bar{I}_{\omega }(\lambda a)}{\bar{K}%
_{\omega }(\lambda a)}K_{\omega }(\lambda \xi )K_{\omega }(\lambda
\xi ^{\prime })\cosh [\omega (\tau -\tau ^{\prime })],  \label{Wb}
\end{equation}%
and is finite in the coincidence limit for $\xi >a$. The
divergences are contained in the first term corresponding to the
Fulling--Rindler vacuum without boundaries. The VEVs of the field
square and the energy-momentum tensor can be found by making use
of the formulae for the Wightman function and relations
(\ref{mphi2VEV}) and (\ref{mTikVEV}). The corresponding results
can be found in Ref. \cite{SahaRind1}.

\subsection{Wightman function in the region between two plates}

Now we consider the vacuum in the region between two plates
situated in the right Rindler wedge and having the coordinates
$\xi =a$ and $\xi =b$, $b>a$. On the surfaces of the plates the
scalar field satisfies Robin boundary conditions (\ref{mrobcond}),
in general, with different coefficients for separate plates. In
terms of the Rindler coordinate $\xi $ these conditions are
written in the form $\left(
\tilde{A}_{j}+n^{(j)}\tilde{B}_{j}\partial /\partial \xi \right)
\varphi =0$, for $\xi =j=a,b$. In the region between the plates,
the eigenfunctions satisfying the boundary condition on the
plate $\xi =b$ have the form%
\begin{equation}
\varphi _{\sigma }(x)=C_{\sigma }G_{i\omega }^{(b)}(\lambda
b,\lambda \xi )e^{i{\bf kx}-i\omega \tau },  \label{R2eigfunc}
\end{equation}%
where the function $G_{\nu }^{(j)}(u,v)$ is defined by formula
(\ref{Geab}). From the boundary condition on the plate $\xi =a$ we
find that the possible values for $\omega $ are roots to the
equation
\begin{equation}
G_{i\omega }^{ab}(\lambda a,\lambda b)=\bar{I}_{i\omega }^{(b)}(\lambda b)%
\bar{K}_{i\omega }^{(a)}(\lambda a)-\bar{K}_{i\omega }^{(b)}(\lambda b)\bar{I%
}_{i\omega }^{(a)}(\lambda a)=0,  \label{Deigfreq}
\end{equation}%
and the barred notations are defined by formula (\ref{barnotab}) with $A_{j}=%
\tilde{A}_{j}$, $B_{j}=n^{(j)}\tilde{B}_{j}/j$. For a fixed
$\lambda $, the equation (\ref{Deigfreq}) has an infinite set of
real solutions with respect to $\omega $. We will denote them by
$\Omega _{n}=\Omega _{n}(\lambda a,\lambda b)$, $\Omega _{n}>0$,
$n=1,2,\ldots $, and will assume that they are arranged in the
ascending order $\Omega _{n}<\Omega _{n+1}$. In addition to the
real zeros, in dependence of the values of the ratios
$A_{j}/B_{j}$, equation (\ref{Deigfreq}) can have a finite set of
purely imaginary solutions. The presence of such solutions leads
to the modes with an imaginary frequency and, hence, to the
unstable vacuum. In the
consideration below we will assume the values of the coefficients in Eq. (%
\ref{mrobcond}) for which the imaginary solutions are absent and
the vacuum is stable.

The coefficient $C_{\sigma }$ in formula (\ref{R2eigfunc}) is
determined from the normalization condition. Taking into account
boundary conditions, for this coefficient one finds
\begin{equation}
C_{\sigma }^{2}=\left. \frac{\left( 2\pi \right)
^{1-D}\bar{I}_{i\omega
}^{(a)}(\lambda a)}{\bar{I}_{i\omega }^{(b)}(\lambda b)\frac{\partial }{%
\partial \omega }G_{i\omega }^{ab}(\lambda a,\lambda b)}\right\vert _{\omega
=\Omega _{n}}.  \label{Dnormc}
\end{equation}%
Now substituting the eigenfunctions into the mode sum formula (\ref%
{mfieldmodesum}), one finds
\begin{equation}
W(x,x^{\prime })=\int d{\bf k}\,\frac{e^{i{\bf k}({\bf x}-{\bf x}^{\prime })}%
}{(2\pi )^{D-1}}\sum_{n=1}^{\infty }\frac{\bar{I}_{i\omega
}^{(a)}(\lambda
a)e^{-i\omega (\tau -\tau ^{\prime })}}{\bar{I}_{i\omega }^{(b)}(\lambda b)%
\frac{\partial }{\partial \omega }G_{i\omega }^{ab}(\lambda a,\lambda b)}%
G_{i\omega }^{(b)}(\lambda b,\lambda \xi )G_{i\omega
}^{(b)}(\lambda b,\lambda \xi ^{\prime })|_{\omega =\Omega _{n}}.
\label{Wigh1}
\end{equation}

To obtain a formula for the summation over $n$, in the GAPF we choose%
\begin{equation}
f(z)=\frac{2i}{\pi }F(z)\sinh \pi z,\quad g(z)=F(z)\frac{\bar{I}%
_{-iz}^{(a)}(\eta )\bar{I}_{iz}^{(b)}(\xi )+\bar{I}_{iz}^{(a)}(\eta )\bar{I}%
_{-iz}^{(b)}(\xi )}{G_{iz}^{ab}(\eta ,\xi )},  \label{fgZi}
\end{equation}%
with a function $F(z)$ analytic in the right half-plane. The
corresponding conditions are satisfied if the function $F(z)$ is
restricted by the
condition%
\begin{equation}
|F(z)|<\epsilon (|z|)e^{-\pi x}\left( \xi /\eta \right)
^{2|y|},\quad z=x+iy,\quad x>0,\quad |z|\rightarrow \infty ,
\label{condFZi}
\end{equation}%
where $|z|\epsilon (|z|)\rightarrow 0$ for $|z|\rightarrow \infty
$. Now
from the GAPF one obtains the summation formula%
\begin{eqnarray}
\sum_{n=1}^{\infty }\frac{\bar{I}_{iz}^{(a)}(\eta )\bar{I}_{-iz}^{(b)}(\xi )%
}{\partial G_{iz}^{ab}(\eta ,\xi )/\partial z}F(z)|_{z=\Omega _{n}} &=&\frac{%
1}{\pi ^{2}}\int_{0}^{\infty }dxF(x)\sinh \pi x  \nonumber \\
&&-\frac{1}{2\pi }\,\int_{0}^{\infty }dz\frac{\bar{I}_{z}^{(a)}(\eta )\bar{I}%
_{-z}^{(b)}(\xi )}{G_{z}^{ab}(\eta ,\xi )}\left[
F(iz)+F(-iz)\right] . \label{sumformZi2}
\end{eqnarray}

For the further evaluation of the Wightman function we apply to
the sum over $n$ in Eq. (\ref{Wigh1}) the summation formula
(\ref{sumformZi2}) taking
\begin{equation}
F(z)=\frac{G_{i\omega }^{(b)}(\lambda b,\lambda \xi )G_{i\omega
}^{(b)}(\lambda b,\lambda \xi ^{\prime })}{\bar{I}_{iz}^{(b)}(\lambda b)\bar{%
I}_{-iz}^{(b)}(\lambda b)}e^{-iz(\tau -\tau ^{\prime })}.
\label{FtoAPF}
\end{equation}%
Condition (\ref{condFZi}) for this function is satisfied if
$a^{2}e^{|\tau -\tau ^{\prime }|}<\xi \xi ^{\prime }$. In
particular, this is the case in the coincidence limit $\tau =\tau
^{\prime }$ in the region under consideration: $\xi ,\xi ^{\prime
}>a$. As a result, for the Wightman
function one obtains the expression%
\begin{eqnarray}
W(x,x^{\prime }) &=&W^{(b)}(x,x^{\prime })-\int \frac{d{\bf k\,}e^{i{\bf k}(%
{\bf x}-{\bf x}^{\prime })}}{\pi (2\pi )^{D-1}}\int_{0}^{\infty
}d\omega
\,\Omega _{b\omega }(\lambda a,\lambda b)  \nonumber \\
&&\times G_{i\omega }^{(b)}(\lambda b,\lambda \xi )G_{i\omega
}^{(b)}(\lambda b,\lambda \xi ^{\prime })\cosh [\omega (\tau -\tau
^{\prime })],  \label{Wigh3}
\end{eqnarray}%
where
\begin{equation}
W^{(b)}(x,x^{\prime })=\int \frac{d{\bf k}e^{i{\bf k}({\bf x}-{\bf x}%
^{\prime })}}{\pi ^{2}(2\pi )^{D-1}}\int_{0}^{\infty }d\omega
\sinh (\pi \omega )e^{-i\omega (\tau -\tau ^{\prime
})}\frac{G_{i\omega }^{(b)}(\lambda
b,\lambda \xi )G_{i\omega }^{(b)}(\lambda b,\lambda \xi ^{\prime })}{\bar{I}%
_{i\omega }^{(b)}(\lambda b)\bar{I}_{-i\omega }^{(b)}(\lambda b)},
\label{Wigh1pl}
\end{equation}%
is the Wightman function in the region $\xi <b$ for a single plate
at $\xi =b $. The latter can be presented in the form
\begin{equation}
W^{(b)}(x,x^{\prime })=W_{{\rm R}}(x,x^{\prime })+\left\langle
\varphi (x)\varphi (x^{\prime })\right\rangle _{{\rm b}},
\label{G+2}
\end{equation}%
where $W_{{\rm R}}(x,x^{\prime })$ is the Wightman function for
the right Rindler wedge without boundaries and
\begin{equation}
\left\langle \varphi (x)\varphi (x^{\prime })\right\rangle _{{\rm
b}}=-\int
\frac{d{\bf k}e^{i{\bf k}({\bf x}-{\bf x}^{\prime })}}{\pi (2\pi )^{D-1}}%
\int_{0}^{\infty }d\omega \frac{\bar{K}_{\omega }^{(b)}(\lambda b)}{\bar{I}%
_{\omega }^{(b)}(\lambda b)}I_{\omega }(\lambda \xi )I_{\omega
}(\lambda \xi ^{\prime })\cosh [\omega (\tau -\tau ^{\prime })]
\label{phi212}
\end{equation}%
is induced in the region $\xi <b$ by the presence of the plate at
$\xi =b$. Note that the representation (\ref{G+2}) with
(\ref{phi212}) is valid under the assumption $\xi \xi ^{\prime
}<b^{2}e^{|\tau -\tau ^{\prime }|}$. The results of the
investigation for the VEVs of the field square and the
energy-momentum tensor, as well as for the vacuum interaction
forces between the plates can be found in Ref. \cite{SahaRind2}.

\section{Other applications}

\label{sec:Applic}

In this section we briefly outline the problems to which the GAPF
is applied in the investigation of the quantum vacuum effects. For
a scalar field obeying the Robin boundary condition on two
parallel plates in the Minkowski
spacetime the corresponding normal modes are the zeros of the function $%
(1-b_{1}b_{2}z^{2})\sin z-(b_{1}+b_{2})z\cos z$, where the coefficients $%
b_{i}$ are related with the coefficients in the boundary
conditions by the formula $b_{i}=B_{i}/A_{i}a$, and $\ a$ is the
distance between the plates. The corresponding summation formula
is obtained and is applied for the evaluation of the vacuum
energy-momentum tensor in Ref. \cite{Rome02}. Similar type of
series arise in the investigation of the Casimir effect for
a scalar field with non-local boundary condition on parallel plates \cite%
{Saha06nonloc}. In this case the coefficients in the Robin
boundary conditions are functions on the quantum numbers
corresponding to the degrees of freedom parallel to the plates.

Series over the zeros of the cylindrical functions arise in the
investigation of quantum vacuum effects in presence of boundaries
with cylindrical symmetry. For the electromagnetic field inside a
cylindrical shell the eigenmodes are zeros of the function
$J_{l}(z)$ for the TM modes
and the zeros of the function $J_{l}^{\prime }(z)$ for the TE modes, $%
l=1,2,\ldots $. The VEVs of the energy-momentum tensor in this
region are
evaluated in Ref. \cite{Saha88} making use of summation formula (\ref{sumJ1}%
). For the electromagnetic vacuum in the region between two
coaxial cylindrical shells with radii $a$ and $b$ the eigenmodes
are zeros of the function
$J_{l}(z)Y_{l}(zb/a)-Y_{l}(z)J_{l}(zb/a)$ for the TM modes and of
the function $J_{l}^{\prime }(z)Y_{l}^{\prime
}(zb/a)-Y_{l}^{\prime }(z)J_{l}^{\prime }(zb/a)$ for the TE modes.
For the investigation of the corresponding VEVs of the
energy-momentum tensor in Ref. \cite{Saha88b} the GAPF is used.
For a scalar field satisfying Robin boundary condition on
cylindrically symmetric boundaries the Wightman function, VEVs for
the field
square and the energy-momentum tensor are investigated in Refs. \cite%
{Rome01,Saha06} for the cases of a single and two coaxial
boundaries by using summation formulae (\ref{sumJ1}) and
(\ref{cor3form}). For the geometry of a wedge with the opening
angle $\phi _{0}$ and with a coaxial cylindrical boundary, the
eigenmodes for a Dirichlet scalar field are zeros of the function
$J_{\pi l/\phi _{0}}(z)$. For the investigation of the vacuum
densities in Refs. \cite{Reza02} the GAPF is applied to the
corresponding mode-sums. The geometry of a cylindrical boundary in
the cosmic string spacetime is considered in Ref. \cite{Beze06b}.

Series over the zeros of combinations of the cylindrical functions
arise in the investigation of the one-loop quantum vacuum effects
in the Randall-Sundrum type braneworlds. The corresponding
background spacetime is described by the line element
$ds^{2}=e^{-2k_{D}y}\eta _{\mu \nu }dx^{\mu
}dx^{\nu }-dy^{2}$, where $\eta _{\mu \nu }$ is the metric for the $D$%
-dimensional Minkowski spacetime and $k_{D}$ is the inverse radius
for the background AdS spacetime. In the two-brane model the
radial part of the eigenfunctions is a combination of the
cylindrical functions. The corresponding eigenmodes are obtained
by imposing the boundary conditions on the branes and are zeroes
of the function
\begin{equation}
\bar{J}_{\nu }^{(a)}(z)\bar{Y}_{\nu
}^{(b)}(ze^{k_{D}(b-a)})-\bar{Y}_{\nu
}^{(a)}(z)\bar{J}_{\nu }^{(b)}(ze^{k_{D}(b-a)}),\;\nu =\sqrt{%
(D/2)^{2}-D(D+1)\zeta +m^{2}/k_{D}^{2}},  \label{cnu}
\end{equation}%
where $a$ and $b$, $a<b$, are the $y$-coordinates of the branes.
The corresponding two-point function, the VEVs of the field square
and the energy-momentum tensor are investigated in Ref.
\cite{Saha05b}. Similar issues for higher dimensional braneworlds
with compact internal spaces are considered in Refs.
\cite{Saha06a}.

\section{Conclusion}

We have considered the application of the GAPF for the
renormalization of the VEVs of local physical observables in
quantum field theory on manifolds with boundaries. These VEVs
contain series over the eigenmodes of the corresponding problem
which are zeroes of the corresponding part in the eigenfunctions.
We have shown that the GAPF allows to obtain the summation
formulae for the series over these zeros, which explicitly extract
from the VEVs the parts corresponding to the bulk without
boundaries. In addition the boundary induced parts are presented
in terms of integrals which are exponentially convergent for the
points away from the boundaries. As a result, the renormalization
is necessary for the boundary-free parts only and this procedure
is the same as that in quantum field theory without boundaries. In
general, the physical quantities in the problems with boundary
conditions can be classified into two main types. For the
quantities of the first type the contribution of the higher modes
into the boundary induced effects is suppressed by the parameters
already present in the idealized model. Examples of such
quantities are the local physical observables for points away from
the boundaries and the interaction forces between disjoint bodies.
As we have seen, the Abel-Plana formula provides an efficient way
to evaluate the quantities of this type. For the quantities from
the second type, such as the energy density on the boundary and
the total vacuum energy, the contribution of the arbitrary higher
modes is dominant and they contain divergences which cannot be
eliminated by the standard renormalization procedure of quantum
field theory without boundaries. In this case an additional
subtractions are necessary. Methods to extract finite results for
the quantities of the second type are developed in Refs.
\cite{Zeta}.

The work was supported by PVE/CAPES program and in part by the
Armenian Ministry of Education and Science Grant No. 0124.

\end{document}